\documentclass[prd,aps,showpacs,preprintnumbers,amsmath,amssymb,superscriptaddress,twocolumn,floatfix]{revtex4}
\usepackage{alltt}
\usepackage{epsfig}
\usepackage{calc,rotating,xspace}
\usepackage{pifont}
\usepackage[mediumspace,mediumqspace,Grey,squaren]{SIunits}
\usepackage{times}
\usepackage{mathptmx}
\newcommand{\met}{\mbox{${\hbox{$p$\kern-0.35em\lower-.1ex\hbox{/}}}_T\:$}}
\newcommand{\vecmet}{\mbox{${\hbox{$\vec{E}$\kern-0.6em\lower-.1ex\hbox{/}}}_T\:$}}

\begin{document}
\title{Comparison of {\sc horace} and {\sc photos} Algorithms for Multi-Photon Emission in the Context of the $W$ Boson Mass Measurement
}


\author{A. V. Kotwal}
\affiliation{
              Physics Department, 
              Duke University, 
              Durham, NC 27708, USA. }
              \email{ashutosh.kotwal@duke.edu}     
\author{           B. Jayatilaka}
\affiliation{              Scientific Computing Division, Fermi National Accelerator Laboratory,
	      Batavia, IL 60510, USA.  
}

\begin{abstract}
The $W$ boson mass measurement is sensitive to QED radiative corrections due to virtual photon loops and real photon emission. The largest shift in the measured mass, which depends on the transverse momentum spectrum of the charged lepton from the boson decay, is caused by the emission of real photons from the final-state lepton. There are a number of calculations and codes available to model the final-state photon emission. We perform a detailed study, comparing the results from the {\sc horace} and {\sc photos} implementations of the final-state multi-photon emission in the context of a direct measurement of the $W$ boson mass at the Tevatron. Mass fits are performed using a simulation of the CDF II detector.
\end{abstract}


\maketitle

%===============================================================================

\section{Introduction}
\label{sec:introduction}
The measurement of the $W$ boson mass ($m_W$) is one of the most interesting precision electroweak observables. In the standard model (SM), the mass of the $W$ boson can be calculated with higher precision~\cite{baak} than the existing measurement uncertainty~\cite{tevcombo,cdfprl,d0prl,d0prd,prdSecond}, thus providing motivation for improving the statistical and systematic uncertainties on the measurement. The comparison between the theoretical prediction and the measurement provides a stringent test of the SM and constrains beyond-standard model (BSM) theories. 

 At hadron colliders, the mass of the $W$ boson is extracted from inclusively-produced $W$ bosons decaying to electrons or muons and the associated neutrinos. At the Tevatron, almost pure samples of such candidate events have been identified with backgrounds typically smaller than 1\%. The momenta of the decay electrons and muons have been measured with a precision of $\sim 0.01$\%, allowing a $W$ boson mass measurement with precision of 0.02\%~\cite{prdSecond}. 

The calibration of the electron and muon momenta is the single most important aspect of the $m_W$ measurement. In the approximation that the $W$ boson undergoes a two-body decay, the distribution of the transverse momentum ($p_T$)~\cite{etaphi} of the charged lepton has the characteristic Jacobian edge at half the mass of $W$ boson.  In practice, electroweak radiative corrections modify the lepton $p_T$ spectrum, mainly due to the emission of photons from the decay lepton. If no correction were applied for this radiative process, the $m_W$ measurement would be biased by $\approx 200$~MeV~\cite{prd}~\footnote{Throughout this paper, we use the convention $\hbar = c = 1$.}. 

 In the first Run II measurement~\cite{prd} of the $W$ boson mass, the {\sc wgrad}~\cite{wgrad} and {\sc zgrad}~\cite{zgrad} programs were used to calculate the QED radiative correction. {\sc wgrad} and {\sc zgrad} are exact next-to-leading order (NLO) electroweak matrix element calculations of the $ q \bar {q} \rightarrow W \gamma \rightarrow l \nu \gamma$ and $ q \bar {q} \rightarrow Z \gamma \rightarrow l \bar{l} \gamma$  processes, respectively. In order to increase the precision of the QED radiative correction for the $W$ boson mass measurement, higher-order calculations were used, as implemented in the {\sc horace}~\cite{CarloniCalame:2003ux, CarloniCalame:2005vc, CarloniCalame:2006zq, CarloniCalame:2007cd, horace, horacePrivate} and {\sc photos}~\cite{photosPaper, photos, photosPaper2} programs. These programs calculate the emission of multiple photons with the appropriate rates, energy and angular distributions.

{\sc photos } uses the exact first-order matrix element of $W$ and $Z$ boson decay for the photon emission kernel.  
For multiphoton radiation, {\sc photos} uses an  iterative solution  for this kernel, developed on the basis of an exact and complete phase space parametrization. This ensures not only resummation of leading-logarithm contributions of higher orders, but the infrared region of the phase space is accurately simulated as well~\cite{photos,photosPaper2}. 
 
{\sc horace}  is a parton-level electroweak Monte Carlo generator for precision
simulations of  charged-current and neutral-current Drell-Yan processes. 
 {\sc horace} uses the full phase space for each radiated photon, and
 there is no ordering of the photons (i.e. in angle or transverse momentum) in multi-photon emission~\cite{horacePrivate}.
Two versions of the {\sc horace} 
 program are available. The {\sc old} version~\cite{CarloniCalame:2003ux,CarloniCalame:2005vc} 
implements a multi-photon emission QED parton shower algorithm for the
simulation of final-state radiation (FSR) in the leading-logarithmic
approximation, without initial-state radiation (ISR) and without
interference between ISR and FSR. In this sense the {\sc old horace}  program is
similar to the {\sc photos} program, which also implements multi-photon FSR.
{\sc old horace}  does not include full one-loop electroweak corrections, but 
it mimics the real radiation matrix element for the description of
the photon radiation in $W$ and $Z$ boson decays, in the leading-logarithmic
  approximation~\cite{horacePrivate}. 
There is also a {\sc new horace} program~\cite{CarloniCalame:2006zq, CarloniCalame:2007cd}, which implements multi-photon
ISR and FSR with interference, and also matches each photon to the exact matrix
element calculation of one-loop electroweak corrections and
single-photon emission~\cite{horacePrivate}. 

 The {\sc photos} program provides a generic interface to any other event generator such that all charged leptons produced by the latter can be passed through the {\sc photos} FSR algorithm. We use this feature as follows. We generate $W$ and $Z$ boson events for the Tevatron $p \bar{p}$ collisions at $\sqrt{s} = 1.96$~TeV,  including higher-order QCD matrix elements and
QCD resummation effects, but without loops or emission of electroweak bosons.  We interface these events to {\sc photos} such that the events from the chain contain  the QED-FSR photons added by {\sc photos}. We save photons with $p_T > 0.4$~MeV  and the events are processed with a detector simulation~\cite{prd,prdSecond}  to make the pseudo-data and the mass-fitting templates. 

 In this paper we present comparisons between the distributions and the mass-fitting results obtained from the {\sc old} {\sc horace} and {\sc photos} programs. 

\section{ Electron channel comparisons }

To make a direct comparisons between quantities sensitive to QED physics, we need to ensure that the underlying boson and lepton distributions are identical between {\sc old} {\sc horace} and {\sc photos}. For this purpose we use the ``Born'' mode of {\sc old} {\sc horace} to generate Born-level $q^\prime \bar{q} \rightarrow W \rightarrow l \nu$ and $ q \bar{q} \rightarrow \gamma^* / Z \rightarrow l \bar{l}$ events, which are then processed through {\sc photos}. The Born mode generates these purely $2 \to  1 \to 2$ parton processes with no radiative photons. These events are compared with events from {\sc old} {\sc horace} run in the QED multi-photon emission FSR mode. Both {\sc old} {\sc horace} and {\sc photos} are run in the ``exponentiation'' mode. All of the events used in these comparisons have unit weights. For all generated events we make a generator-level cut on the partonic center-of-mass energy $\sqrt {\hat{s}} > 40$~GeV to remove the contribution of the photon pole for neutral-current events. For consistency, we also apply this cut on the charged-current events. 

\begin{figure}[ht]
\begin{center}
\includegraphics[width=1.05\columnwidth]{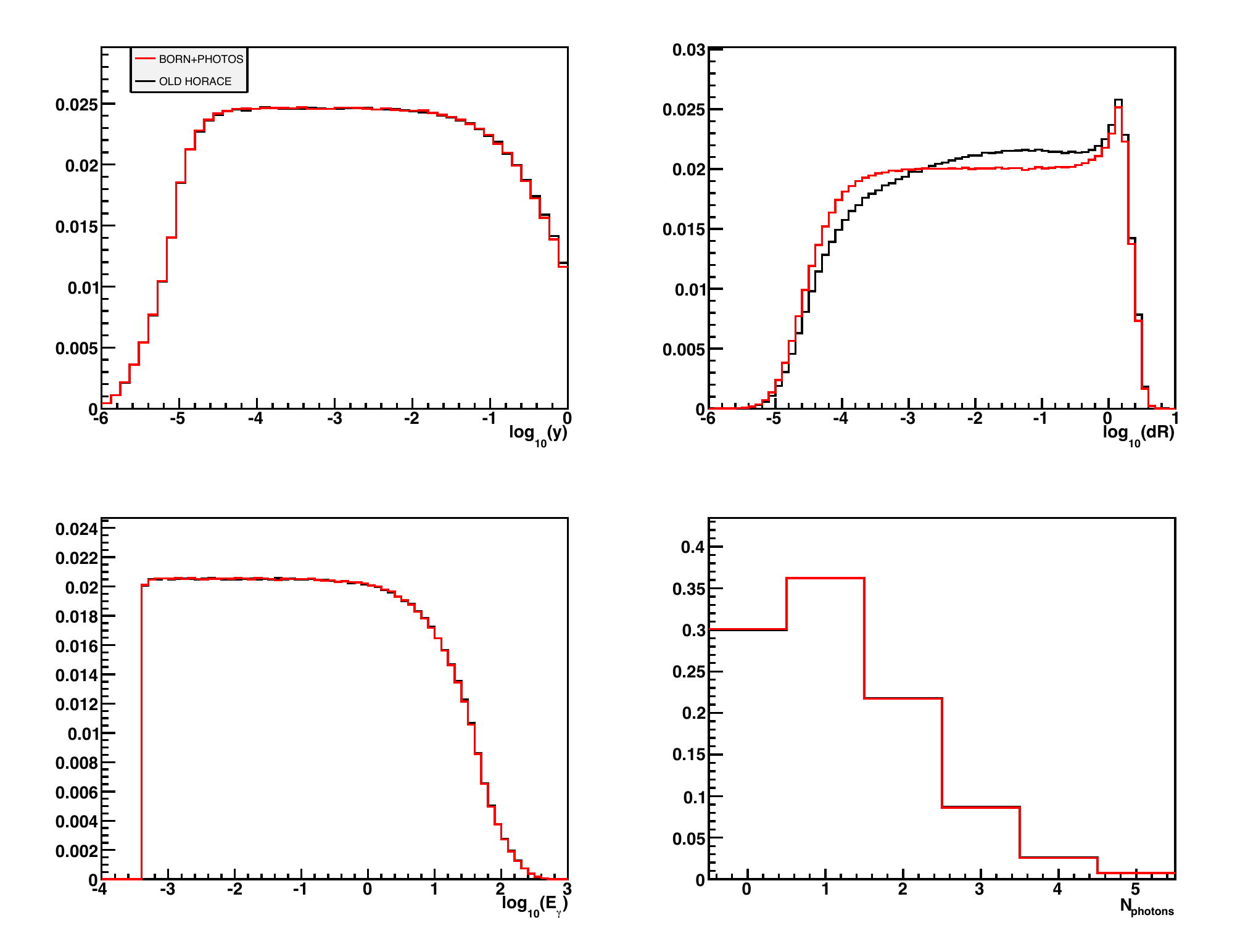}
\end{center}
\caption{Clockwise from top-left: comparisons of the distributions of $\log_{10}(y_\gamma)$, $\log_{10}(\Delta R (l \gamma))$, $n_\gamma$, and $\log_{10}(E_\gamma/$GeV) for the $\gamma^*/Z \rightarrow e^+ e^- + n\gamma$ process between the ``Born'' mode of {\sc old} {\sc horace} interfaced with {\sc photos} and {\sc old} {\sc horace} in the exponentiation mode. The smaller of the two $\Delta R$ values with respect to the two electrons is shown.  }
\label{zeeDistributions}
\end{figure}

 In Fig.~\ref{zeeDistributions} we compare the distributions for photon emission rates as well as the energy and angular distributions for the $\gamma^*/Z \rightarrow e^+ e^- + n\gamma$ process. For these comparisons we consider photons with energy $E_\gamma > 0.4$ MeV; photons with lower energy than this threshold are not counted and ignored in the distributions. In addition to the number $n_\gamma$ of photons
 emitted, we find that the distributions of the following quantities are useful to compare: $\log_{10} (E_\gamma /$GeV), the fractional photon energy $y_\gamma \equiv E_\gamma / (E_\gamma + E_l)$ (where $E_l$ is the energy of the final-state lepton) and $\Delta R ( l \gamma) \equiv \sqrt {(\Delta \eta_{l \gamma})^2 
 + (\Delta \phi_{l \gamma})^2 }$ (the $\eta-\phi$ angular separation between a photon and the final-state lepton~\cite{etaphi}).   
 
\begin{figure}[h]
\begin{center}
\includegraphics[width=1.05\columnwidth]{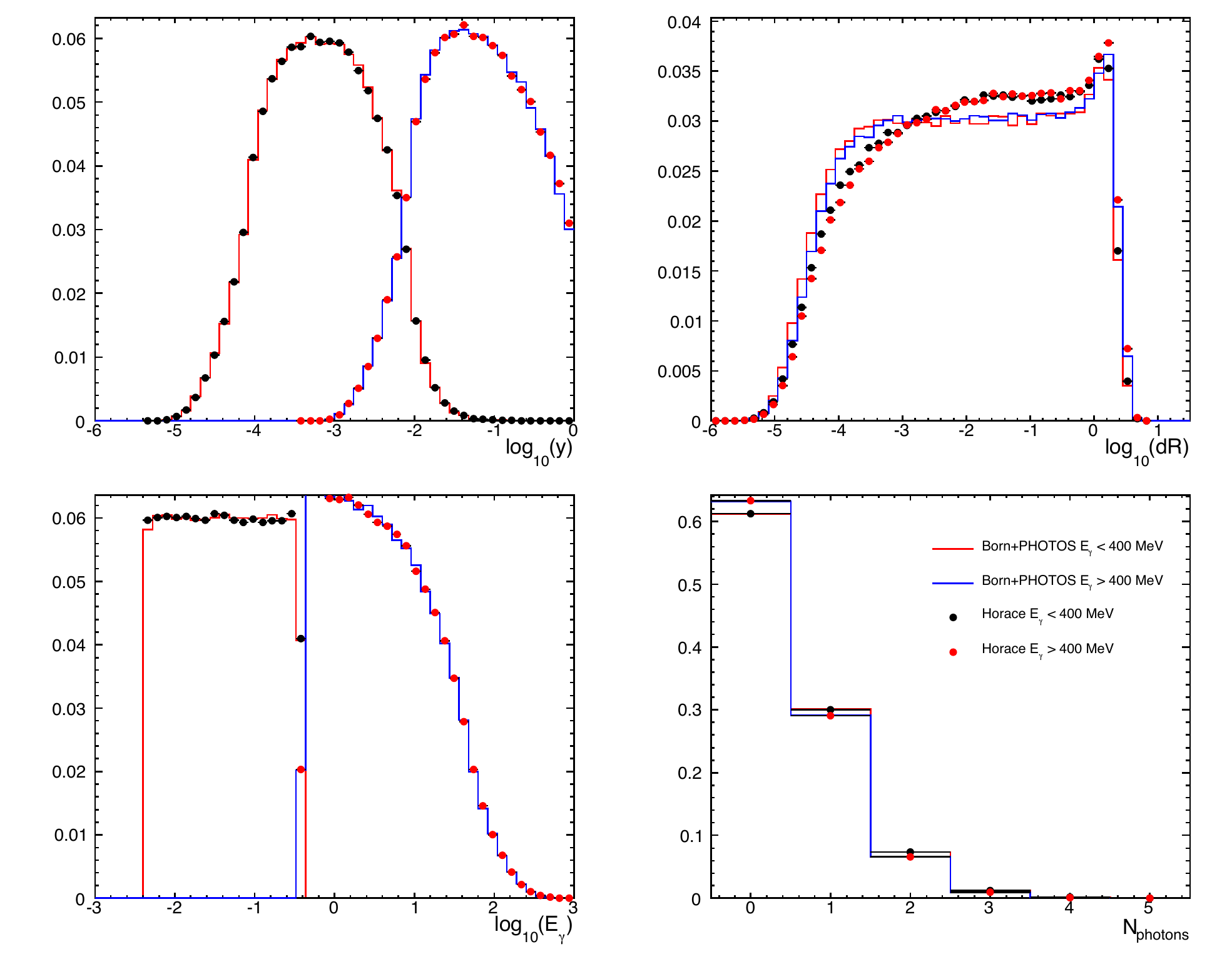}
\end{center}
\caption{Clockwise from top-left: comparisons of the distributions of $\log_{10}(y_\gamma)$, $\log_{10}(\Delta R (l \gamma))$, $n_\gamma$, and $\log_{10}(E_\gamma/$GeV) for the $\gamma^*/Z \rightarrow e^ + e^- + n\gamma$ process  between the ``Born'' mode of {\sc old} {\sc horace} interfaced with {\sc photos} and {\sc old} {\sc horace}  in the exponentiation mode. The comparisons are shown separately  for low ($E_\gamma < 400$~MeV) and high-energy  ($E_\gamma > 400$~MeV) photons. The smaller of the two
 $\Delta R$ values with respect to the two electrons is shown.  }
\label{zeeDistributionsEcut}
\end{figure}

\begin{figure}[h]
\begin{center}
\includegraphics[width=1.05\columnwidth]{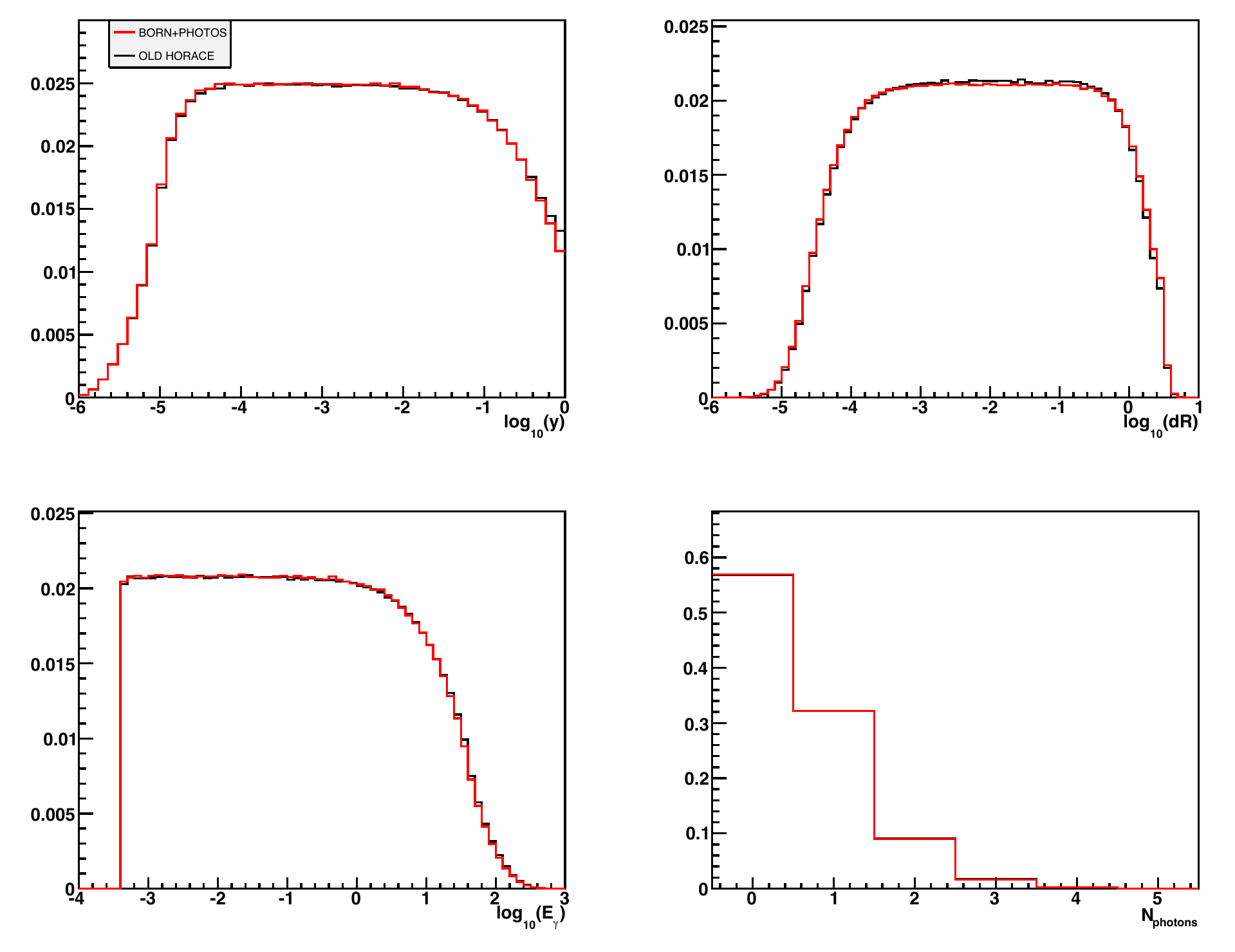}
\end{center}
\caption{Clockwise from top-left: comparisons of the distributions of $\log_{10}(y_\gamma)$, $\log_{10}(\Delta R (l \gamma))$, $n_\gamma$, and $\log_{10}(E_\gamma/$GeV) for the $W^+ \rightarrow e^+ \nu + n\gamma$ process between the ``Born'' mode of {\sc old} {\sc horace} interfaced with {\sc photos} and {\sc old} {\sc horace}  in the exponentiation mode. The $\Delta R$ is computed with respect to the positron. }
\label{wenuDistributions}
\end{figure}

Figure~\ref{zeeDistributionsEcut} shows the photon distributions separately for low-energy and high-energy photons. These distributions show that the angular distribution is almost independent of the photon energy, allowing us to draw conclusions from the inclusive photon distributions. 

The comparisons between {\sc old horace} and {\sc photos} show good agreement in the photon emission rates and the photon energy distributions. For the $\gamma^*/Z \rightarrow e^+ e^- + n \gamma$ process, the photon angular distribution shows about 10\% difference at small angles, a difference that is not correlated with photon energy. As we show in Sec.~\ref{massFits}, this difference does not cause a relative shift in the fitted $Z \to ee$ mass between the two algorithms.

\begin{figure}[h]
\begin{center}
\includegraphics[width=1.05\columnwidth]{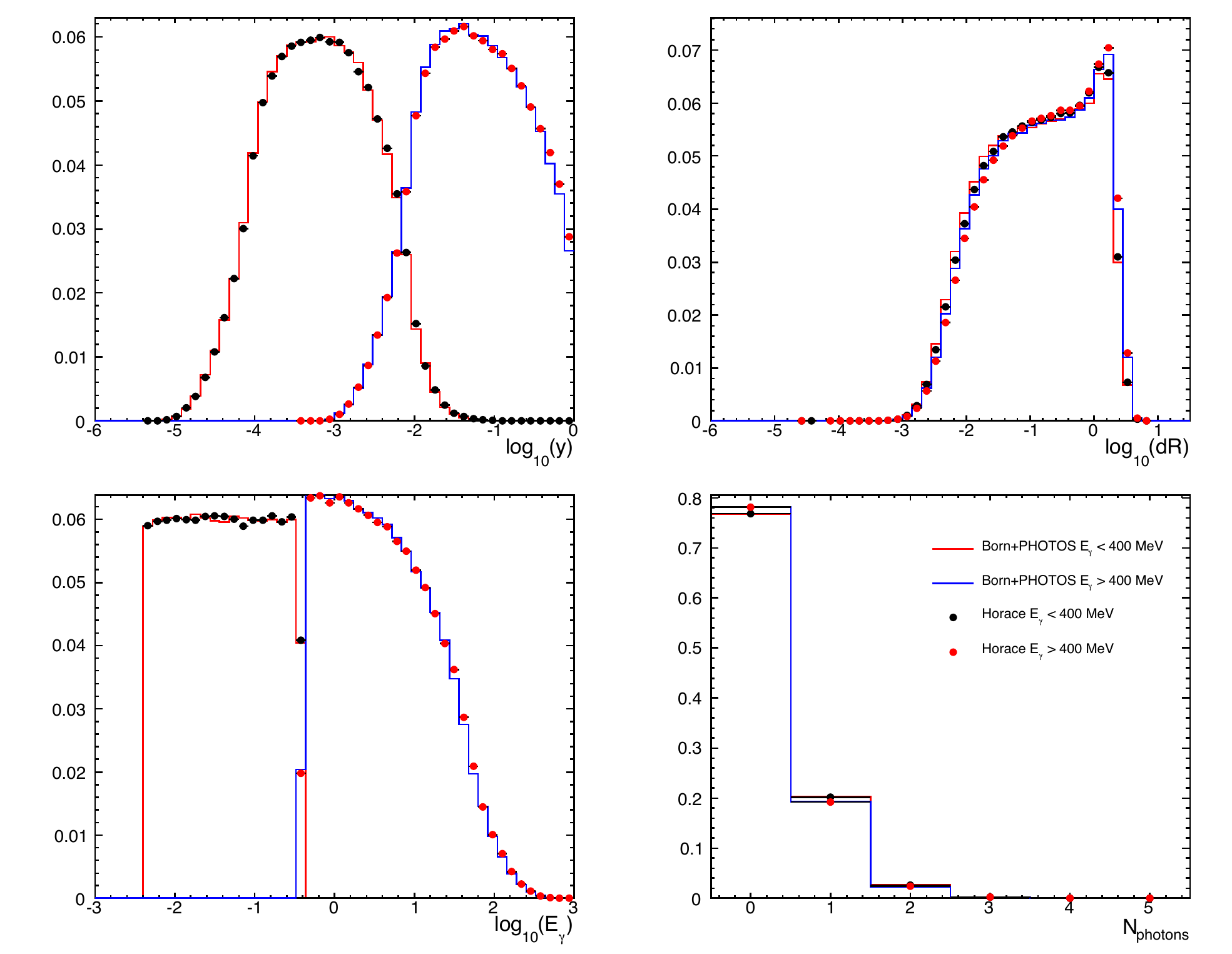}
\end{center}
\caption{Clockwise from top-left: comparisons of the distributions of $\log_{10}(y_\gamma)$, $\log_{10}(\Delta R (l \gamma))$, $n_\gamma$, and $\log_{10}(E_\gamma/$GeV) for the $\gamma^*/Z \rightarrow \mu^ + \mu^- + n\gamma$ process, separated into low ($E_\gamma < 400$~MeV) and high-energy ($E_\gamma > 400$~MeV) photons. The smaller of the two $\Delta R$ values with respect to the two muons is shown.  }
\label{zmmDistributionsEcut}
\end{figure}

\section{ Muon channel comparisons }

\begin{figure}[h]
\begin{center}
\includegraphics[width=1.05\columnwidth]{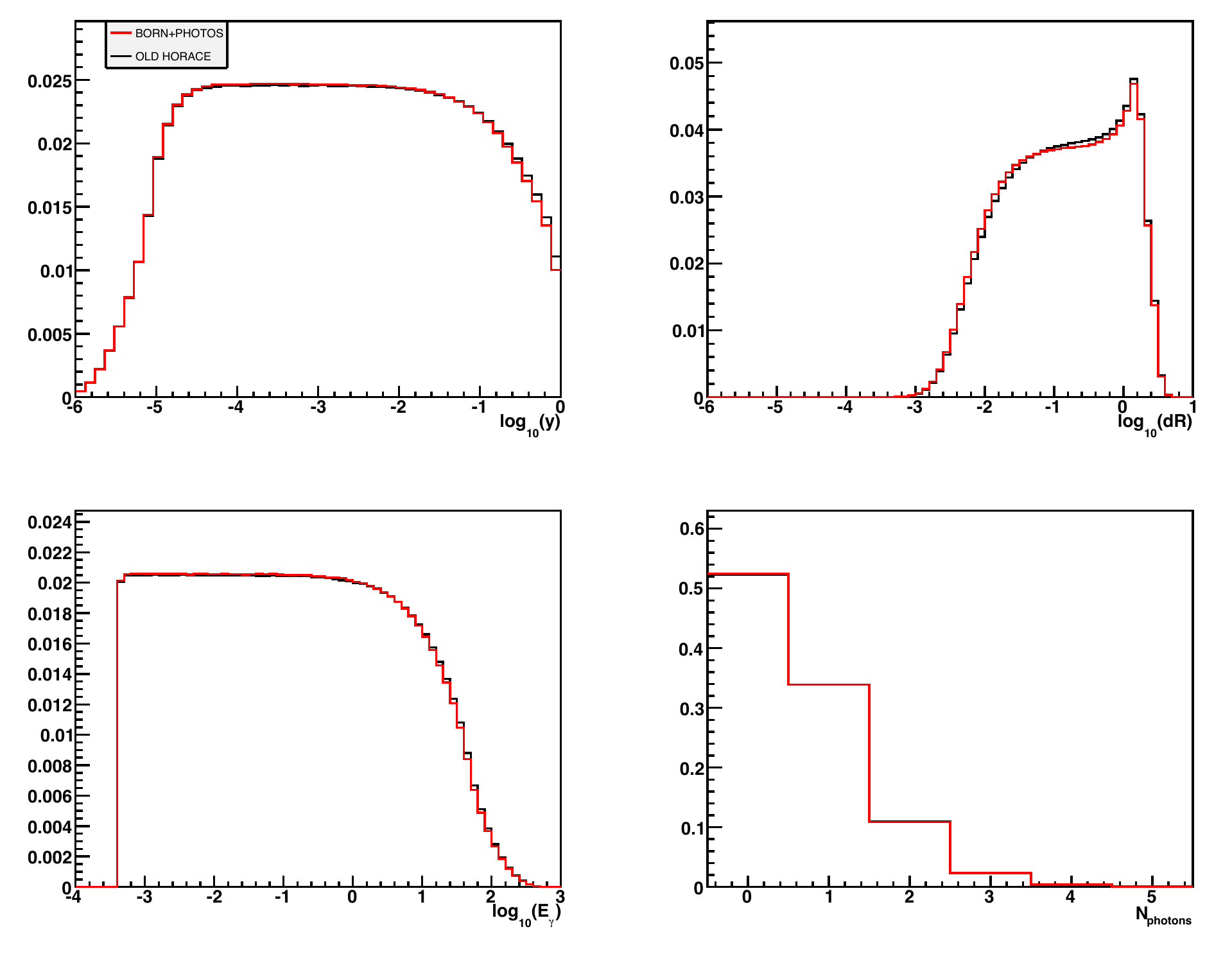}
\end{center}
\caption{Clockwise from top-left: comparisons of the distributions of $\log_{10}(y_\gamma)$, $\log_{10}(\Delta R (l \gamma))$, $n_\gamma$, and $\log_{10}(E_\gamma/$GeV) for the $\gamma^*/Z \rightarrow \mu^+ \mu^- + n\gamma$ process  between the ``Born'' mode of {\sc old} {\sc horace} interfaced with {\sc photos} and {\sc old} {\sc horace}  in the exponentiation mode. The smaller of the two $\Delta R$ values with respect to the two muons is shown.   }
\label{zmmDistributions}
\end{figure}

\begin{figure}[h]
\begin{center}
\includegraphics[width=1.05\columnwidth]{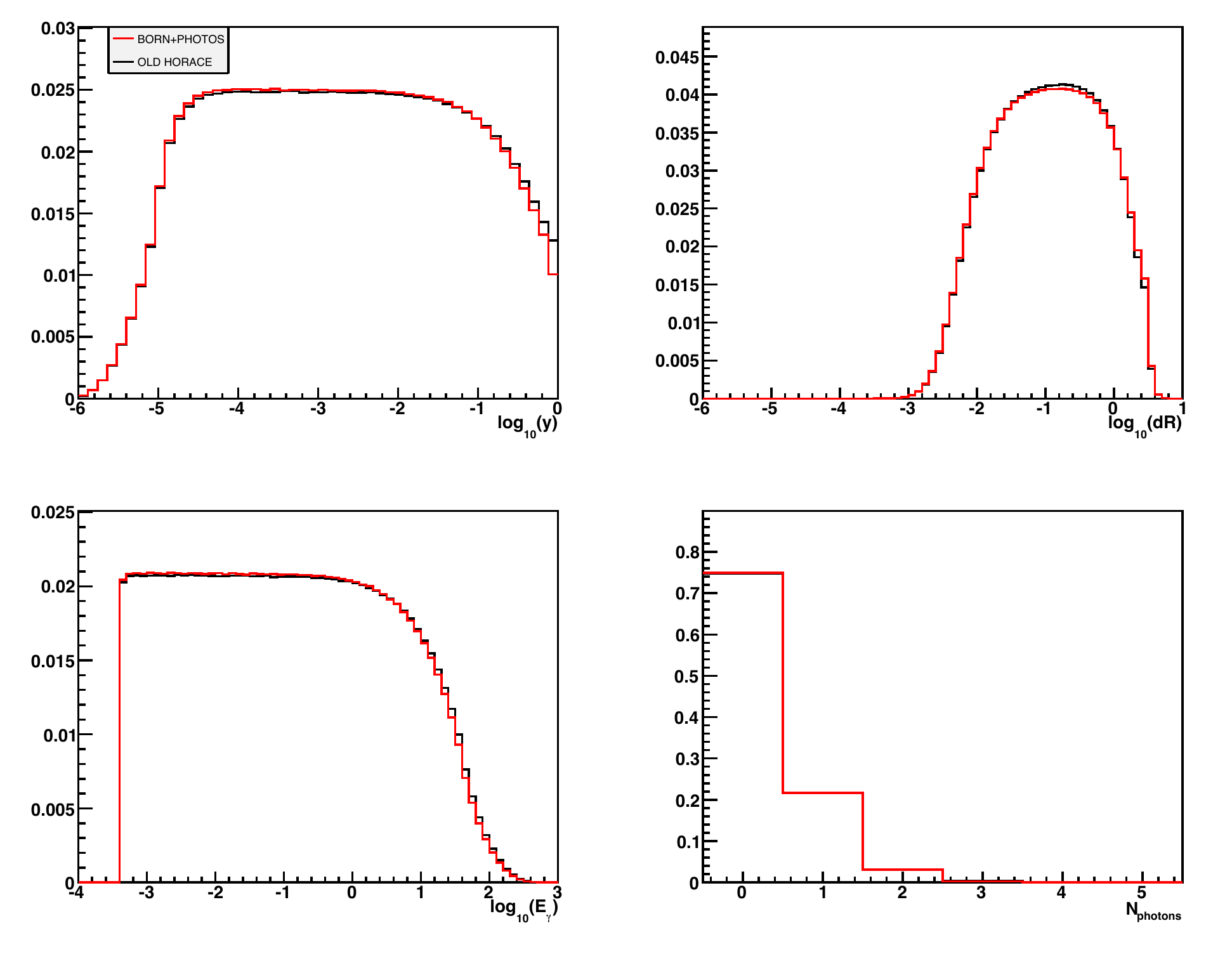}
\end{center}
\caption{Clockwise from top-left: comparisons of the distributions of $\log_{10}(y_\gamma)$, $\log_{10}(\Delta R (l \gamma))$, $n_\gamma$, and $\log_{10}(E_\gamma/$GeV)  for the $W^+ \rightarrow \mu^+ \nu + n\gamma$ process between the ``Born'' mode of {\sc old} {\sc horace} interfaced with {\sc photos} and {\sc old} {\sc horace}  in the exponentiation mode. The $\Delta R$ is computed with respect to the muon.   }
\label{wmunuDistributions}
\end{figure}

We repeat the above comparisons for the muon channel.  In Figs.~\ref{zmmDistributions} and~\ref{wmunuDistributions} we compare the distributions for photon emission rates as well as the energy and angular distributions for the $\gamma^*/Z \rightarrow \mu^+  \mu^- + n\gamma$ and  $W^+ \rightarrow \mu^+ \nu + n\gamma$ processes respectively. The rates and distributions are in good agreement between {\sc old horace} and {\sc photos}. 

\section{Mass Fits}
\label{massFits}
We quantify the impact of the small differences in photon rates and distributions between {\sc old horace} and {\sc photos} in terms of shifts in the fitted $W$ boson masses. For this purpose,  
we propagate high-statistics {\sc old horace} and {\sc photos} samples through the parameterized CDF detector simulation used in the $W$ boson mass measurement~\cite{prdSecond}. We generate templates and pseudo-data from both samples and perform the mass fits to these pseudo-data using these templates, in the same manner that templates are used to fit the collider data. 

 We perform fits to the distributions of transverse quantities in $W$ boson events; the charged lepton $p_T$, neutrino $p_T$, and transverse mass $m_T$~\footnote{The transverse mass is defined as  $m_T=\sqrt{2p_T^\ell p_T^{~\nu}(1 - \cos\Delta\phi)}$, where $\Delta\phi$ is the azimuthal angle between the charged lepton and neutrino momenta in the transverse plane.}.  In the electron channel, we also perform a fit to the ratio of electron calorimeter energy to track momentum ($E/p$), which is used by the CDF experiment to obtain the calorimeter calibration using the electron track. In $Z$ boson events, we fit the distributions of the $Z$-boson invariant-mass obtained from electron calorimeter deposition measurements (cluster mass) and from track momentum measurements of electrons and muons (track mass). 

We obtain the difference between the mass fits to the {\sc horace} pseudo-data and the {\sc photos} pseudo-data which quantifies the relevant differences between the two QED codes. Tables~\ref{massDiffsPhotos} and~\ref{massDiffsHorace} shows the differences along with their statistical uncertainties. Table~\ref{massDiffsPhotos} uses {\sc photos} templates and is essentially identical to Table~\ref{massDiffsHorace} which uses {\sc horace} templates. The two tables provide validation that the comparison of the two pseudo-data samples does not depend on the template choice, as long as the same templates are used for the fits being compared. 

\begin{table}
\begin{center}
\begin{ruledtabular}
\begin{tabular}{lcc}
fit type & \multicolumn{2}{c}{$m_{\rm horace} - m_{\rm photos}$ (MeV)}  \\
         & electron & muon \\
\hline
$W$ transverse mass & $0.0 \pm 0.6$ & $0.0 \pm 0.4$ \\
$W$ lepton $p_T$    & $-0.4 \pm 0.4$   & $0.0 \pm 0.4$ \\
$W$ neutrino $p_T$ & $0.6 \pm 0.8$ & $1.4 \pm 0.6$ \\
$W$ $E/p$  & $0.4 \pm 0.1$ & - \\
$Z$ cluster mass & $0.2 \pm 0.4$ & - \\
$Z$ track mass & $-1.0 \pm 0.6$ & $-0.8 \pm 0.3$ \\
\end{tabular}
\end{ruledtabular}
\caption{Difference between {\sc horace} and {\sc photos} pseudo-data in fitted masses. The shift in the dimensionless $E/p$ value has been multiplied by 80~GeV to convert to the equivalent
 shift in the fitted $W$-boson mass.  Templates were made using {\sc photos}. The statistical errors are shown. The templates and pseudo-data use 10 billion events at the generator level as the input to the detector simulation. }
\label{massDiffsPhotos}
\hspace*{1cm}
\end{center}
\end{table}

\begin{table}
\begin{center}
\begin{ruledtabular}
\begin{tabular}{lcc}
fit type & \multicolumn{2}{c}{$m_{\rm horace} - m_{\rm photos}$ (MeV)}  \\
         & electron & muon \\
\hline
$W$ transverse mass & $0.0 \pm 0.6$ & $0.2 \pm 0.4$ \\
$W$ lepton $p_T$    & $-0.6 \pm 0.4$  &  $0.0 \pm 0.4$ \\
$W$ neutrino $p_T$ & $0.6 \pm 0.8$ & $1.6 \pm 0.6$ \\
$W$ $E/p$   & $0.4 \pm 0.1$ & - \\
$Z$ cluster mass & $0.0 \pm 0.4 $ & - \\
$Z$ track mass &  $-1.2 \pm 0.7$ & $-0.8 \pm 0.3$ \\
\end{tabular}
\end{ruledtabular}
\caption{Difference between {\sc horace} and {\sc photos} pseudo-data in fitted masses. Templates were made using {\sc horace}. The shift in the dimensionless $E/p$ value  has been multiplied by  80~GeV to convert to the equivalent  shift in the fitted $W$-boson mass. The statistical errors are shown. The templates and pseudo-data use 10 billion  events at the generator level as the input to the detector simulation. }
\label{massDiffsHorace}
\hspace*{1cm}
\end{center}
\end{table}

\section{Conclusions}
 We find that the QED generators {\sc old horace} and {\sc photos} agree with each other in the photon rates and distributions. The only noticeable difference is in the photon angular distribution for the $\gamma^*/Z \rightarrow e^+ e^- + n\gamma$ process, at small angular separation from the nearest lepton. We quantify the comparison by computing relative $W$ and $Z$-boson mass shifts, and find them to be consistent with $\approx 0.7$~MeV within  statistical uncertainties.  We conclude that a systematic uncertainty of 0.7 MeV would account for any differences in the 
 FSR multi-photon emission between the {\sc horace } and {\sc photos} algorithms. 

\begin{acknowledgments}                                                                                                                                                                         
We wish to thank Ilija Bizjak for his assistance with  the {\sc horace} program, and Zbigniew Was for providing the interface to the {\sc photos} program.
  We wish to thank  
William Ashmanskas, 
Franco Bedeschi, 
Daniel Beecher, 
Ilija Bizjak, 
Kenichi Hatakeyama, 
 Christopher Hays, 
Mark Lancaster, 
Sarah Malik, 
Larry Nodulman, 
Peter Renton, 
Tom Riddick, 
Ravi Shekhar, 
Melvyn Shochet, 
Oliver Stelzer-Chilton, 
Siyuan Sun, 
David Waters,
Yu Zeng, 
  and other  colleagues in the CDF Collaboration for helpful discussions. We also thank Carlo Carloni Calame, Guido Montagna, Alessandro Vicini, Doreen Wackeroth and Zbigniew Was for discussions regarding  electroweak radiative corrections. We acknowledge the support of the U.S. Department of
Energy, Office of High Energy Physics  and the Fermi National Accelerator Laboratory (Fermilab). The computational resources used in this study were provided by Fermilab. 
Fermilab is operated by Fermi Research Alliance, LLC under Contract No. DE-AC02-07CH11359 with the United States Department of Energy.
\end{acknowledgments}


\begin{thebibliography}{99}
\bibitem{etaphi} The pseudorapidity $\eta = - \log [\tan (\theta /2)]$,  where $\theta$ is the polar angle with respect to the beam axis, and  $\phi$ 
 is the azimuthal angle about the beam axis. The transverse momentum $p_T$ is defined as the component of the momentum perpendicular to the beam axis.
\end{thebibliography}
\end{document}